\documentclass[graybox]{svmult}

\usepackage{mathptmx} 
\usepackage{helvet} 
\usepackage{courier} 
\usepackage{type1cm} 
\usepackage{makeidx} 
\usepackage{graphicx} 
\usepackage{gensymb}
\usepackage{amssymb}
\usepackage{amsmath}
\usepackage{bm}
\graphicspath{{Figures/}}
\usepackage{multicol} 
\usepackage[bottom]{footmisc}

\usepackage{xcolor}

\makeindex 

\begin{document}
\title*{Atomistic Modelling of Energy Dissipation in Nanoscale Gears}
\author{Huang-Hsiang Lin, Alexander Croy, Rafael Gutierrez and Gianaurelio Cuniberti}
\authorrunning{H-H Lin, A Croy, R Gutierrez, G Cuniberti}
\institute{Huang-Hsiang Lin \and Alexander Croy \and Rafael Gutierrez \and Gianaurelio Cuniberti \at Institute for Materials Science and Max Bergmann Center of Biomaterials, TU Dresden, 01069 Dresden, Germany,
\email{rafael.gutierrez@tu-dresden.de}}

\maketitle

\abstract{
{Molecule- and solid-state gears build the elementary constituents of nanoscale mechanical machineries. Recent experimental advances in fabrication technologies in the field have strongly contributed to better delineate the roadmap towards the ultimate goal of engineering molecular-scale mechanical devices. To complement experimental studies, computer simulations play an invaluable role, since they allow to address, with atomistic resolution, various fundamental issues such as the transmission of angular momentum in nanoscale gear trains and the mechanisms of energy dissipation at such length scales. We review in this chapter our work addressing the latter problem. Our computational approach is based on classical atomistic Molecular Dynamics simulations. Two basic problems are discussed: (i) the dominant energy dissipation channels of a rotating solid-state nanogear adsorbed on a surface, and (ii) the transmission of rotational motion and frictional processes in a heterogeneous gear pair consisting of a graphene nanodisk and a molecular-scale gear. }}


\section{\label{sec:Introduction}Introduction}

The increasing demands on the miniaturization of digital and analog devices has led to a proliferation of studies aiming at designing efficient strategies to reduce physical dimensions of the device's building blocks while still preserving their basic functionalities. While such approaches have been extensively developed and refined in the domain of electronic devices $-$following both top-down and bottom-up strategies$-$ less work has been invested in the investigation of truly nanoscale mechanical systems with basic components consisting of molecular\cite{JuYun2007, Yang2014,Mailly2020} or solid-state nanogears\cite{Joachim2020}. The most recent advances in fabrication techniques in this field, including bottom-up approaches\cite{Gisbert2019,Soong2000} as well as top-down methods such as focused ion beams\cite{JuYun2007} or electron beams\cite{Deng2011,Yang2014} point, however, in a very promising direction. In this context, there are several general issues, which play an important role in determining the feasibility of a roadmap leading to the fabrication of nanoscale analog devices based on gear trains: (1) the capability of transmitting angular momentum along gear trains, (2) the energy dissipation mechanisms from gear systems into the environment, (3) the atomistic features of the interface between solid-state nanogears and molecule gears, (4) the possible improvement of angular momentum transmission in nanoscale gear trains mediated by lubricants, and (5) the triggering of a rotation at the level of single molecule gears.

In this Chapter we will focus on different aspects related to items (2) and (3) using representative nanogear systems. Studies related to item (1) in molecule gear trains have been previously presented in e.g.\ \cite{Lin2020,PhysRevApplied.13.034024,Hove2018a,Chen2018,Hove2018}. The influence of lubricants in nanoscale solid-state gears has been recently studied in \cite{lin2021effect,doi:10.1021/acs.jpcc.1c04239}. Work related to item (5) is, on the contrary, scarce, see e.g.\ Refs.\ \cite{PES,Lin2019}.

Concerning item (2), it is well-established that the friction laws known from macroscopic systems cannot, in general, be extrapolated to nanoscale systems\cite{Vanossi2013,Gnecco2006,Perssson1995,Wen2017,Manini2016,Sankar2020}, and this has lead to a large number of experimental\cite{Lodge2016,Smith1996,Walker2012,Binnig1987,Balakrishna2014} and model-based\cite{Tomlinson1929,Borner,Panizon2018,Guerra2010} investigations. In general terms, two qualitatively different friction regimes have been identified, which depend on the center-of-mass velocity: (a) a low-speed, stochastic regime ($v\ll $ 1000 cm/s)\cite{Tomassone1997a}, where the system is close to thermal equilibrium and friction is governed by Brownian motion, (b) a high-speed, viscous regime\cite{Tomassone1997a,Guerra2010}, where the system is out of equilibrium and friction is proportional to the number of collisions with surface corrugations per unit time\cite{Guerra2010}. However, in the case of rotational motion, it is not immediately obvious if a clear separation of stochastic and viscous regimes is possible, since high tangential velocity (outer) and low tangential velocity (inner) atoms in a nanoscale gear are present at the same time and the motion of both sets is correlated. Additionally, for molecule gears as well as nanoscale solid-state gears, energy dissipation into internal degrees of freedom due to the larger conformational flexibility of the gears (compared with their macroscopic counterparts) can play a non-negligible role. 
%
For the development of truly nanoscale mechanical machines it is also of interest to address not only the transmission of rotational motion between gears of the same type (either solid-state or molecule gears), but also the interaction of gears with different sizes and different materials (item (3) above). A typical situation would be that of a solid-state nanogear coupled to a molecule gear. The transmission of rotational motion in such a situation is by far not obvious: molecule gears are rather flexible and energy can be dissipated into internal degrees of freedom. Moreover, the contact interface of both gears will have atomic resolution and, therefore, may require a very careful design to ensure good interlocking. Still, recent progress in the miniaturization of solid states gears down to the nanoscale\cite{JuYun2007,Yang2014,Mailly2020,Deng2011} is calling for the study of gear trains and of the transfer of rotation over different length scales. We may have been thus reaching the experimental stage where a single molecule gear (with diameter $\sim$ 1 nm)\cite{Manzano2009} can be rotated by a solid state nanogear (diameter $\sim$ 10 nm)\cite{Yang2014,Mailly2020}. Rotation of single molecule gears\cite{Chiaravalloti2007,Moresco2015,Mishra2015,Perera2013,Manzano2009} or short molecule gear trains\cite{AuYeung2020,WeiHyo2019,Zhang2016} has already been experimentally demonstrated. Building nanoscale devices where the transmission of angular momentum across gears with different sizes becomes feasible, poses many challenges to the experimentalist\cite{Joachim2020}: achieving gear thickness compatibility, the surface preservation with atomistic resolution after transferring the solid state nanogear on it, the mechanical stability of the rotational axle, and last but not least the triggering of the rotational motion and its transmission to gears of different size and composition. 

This chapter is organized as follows: in Sec.\,\ref{sec:Formalism}, we describe briefly the basic Molecular Dynamics simulation details for the three different model systems we are going to discuss in Sec.\,\ref{sec:results}. The latter includes, whenever necessary, more specific information on the simulation details. In Subsec.\,\ref{sec:gearonsubs}, we study the energy dissipation mechanism in a setup consisting of a nanoscale solid-state gear adsorbed on different substrates. In Subsec.\,\ref{sec:graphene} the rotational transmission between a graphene nanodisk and a molecule gear deposited on a Cu surface is addressed, together with an estimate of the vibrational and electronic contributions to rotational friction. The chapter ends with a summary of the results and an outlook in Sec.\,\ref{sec:Conclusion}.

\section{\label{sec:Formalism} Simulation methodology }

For all the simulations presented here we have used the Large-scale Atomic/Molecular Massively Parallel Simulator (LAMMPS) software\cite{Plimpton1995}. The choice of the corresponding classical force fields is, however, dependent on the specific system setup and is briefly described in the following:
\begin{enumerate} 
\item For the solid-state diamond gear on SiO$_2$ substrates (Sec.\ \ref{sec:gearonsubs}), we use the Tersoff potential \cite{Tersoff1988} in the SiO$_2$ bulk, whereas the adaptive intermolecular reactive empirical bond order (AIREBO) potential \cite{Stuart2013} was chosen for the carbon subsystem (gear). The coupling between both subsystems was assumed to be a non-bonding 12-6 Lennard-Jones type of potential. The corresponding parameters are given in Ref.\ \cite{Lin2020a}. When using a graphene substrate for comparison, the AIREBO potential was also used for the surface.
\item The interaction between a graphene nanodisk and a molecule gear on a Cu substrate ((Sec.\ \ref{sec:graphene}) was treated via the Reactive Force Field (ReaxFF)\cite{Plimpton1995}. The upper two Cu layers were coupled to a Nos\'e–Hoover thermostat\cite{Nose1984,Hoover1985} to fix a generic temperature $T$ = 10K, which is of similar order as that from low-temperature STM experiments. To define corresponding collective rotational variables out of the gear atomistic conformations we exploited the nearly rigid-body approximation (NRBA), see Refs.\ \cite{Lin2019a, Lin2020} for technical details.
\end{enumerate}


\section{\label{sec:results} Results}
\subsection{\label{sec:gearonsubs} Diamond solid-state gear on substrates}

In this section we will investigate a solid-state diamond nanogear physisorbed on different substrates. Our aim is to elucidate the dependence of the energy dissipation via friction on the type of substrate. Using classical MD simulations allow us to reach the relevant time scales in the range of $100$ps up to $1$ns \cite{Persson1996}. In the first case, we consider a system consisting of a diamond-based solid state gear (radius $r=5$ nm and thickness $9.8$\AA{}) with involute profile\cite{Norton2010} on top of an $\alpha$-cristobalite SiO$_2$(001) surface. The size of the substrate is 23.0 nm$\times$15.3 nm$\times$1.39 nm and we use periodic boundary conditions in $x$ and $y$ directions. The initial vertical separation between gear and substrate is 2.5\AA{} and, for preventing lateral and vertical motion the bottom layer of the substrate was fixed while the remaining part was allowed to structurally relax. For the sake of comparison, we have also used two additional substrates: amorphous SiO$_2$ and graphene. 

\begin{figure}[t]\centering
 \includegraphics[width=0.8\textwidth]{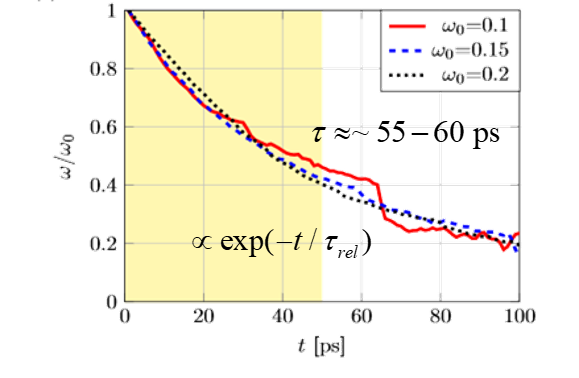}
 \caption{The normalized angular velocity relaxation of a $5$ nm diamond gear on top of $\alpha$-cristobalite SiO$_2$. The initial angular velocities were $\omega=0.1$, $0.15$ and $0.2$ rad/ps. The highlighted yellow region (from $0$ to $50$ ps) was used for extracting the velocity relaxation time $\tau$ by fitting to an exponential function $\propto e^{-t/\tau}$.}
 \label{Fig:viscous}
\end{figure}

\textit{Viscous dissipation limit.$-$} We first address the typical time scales associated with the friction process. Taking into account that the atoms within a rotating gear have large tangential velocities, it is reasonable to expect that we are in the same frictional regime as for linear friction with high translation velocity\cite{Guerra2010}. Therefore, ignoring any applied external torque, the gear dynamics in the high friction limit will be described by a first order differential equation for the angular velocity $\omega$:
\begin{equation}
 I\dot{\omega}+\gamma \omega=0\;,
\end{equation}
where $\gamma$ is the damping coefficient and $I$ is the gear moment of inertia. From this equation we immediately obtain an exponential decay of the velocity with relaxation time given by $\tau_{rel}=\gamma/I$. In the frame of our MD simulations, we give an initial angular velocity to the gear and monitor its decay in time, expecting to find the exponential behavior obtained from Eq.~(1), which in this case would result from the van-der-Waals interactions between the gear and the substrate at a given temperature. The results are displayed in Fig.\ \ref{Fig:viscous} for different initial angular velocities $\omega_0 = 0.1$, $0.15$ and $0.2$ rad/ps and for an MD run of $100$ ps. Notice that in the figure, we have plotted $\omega(t)$ scaled with the corresponding value of $\omega_0$. It becomes apparent that the relaxation process does not sensitively depend on $\omega_0$ and it displays (up to small fluctuations for small initial angular velocity) the expected exponential decay as a function of time. Fitting the obtained curves in the interval from $0$ to $50$ ps (highlighted in yellow in the figure) to a single exponential allows to obtain the corresponding relaxation times $\tau_{rel}=59$, $55$ and $56$ ps for $\omega_0=0.1$, $0.15$ and $0.2$ rad/ps, respectively. Another interesting result is that for the used gear radius of 5 nm the dynamics is that of a deterministic rotor rather than a Brownian rotor, since stochastic fluctuations are increasingly suppressed with increasing gear size. 

\begin{figure}[t!]
\centering
 \includegraphics[width=0.99\textwidth]{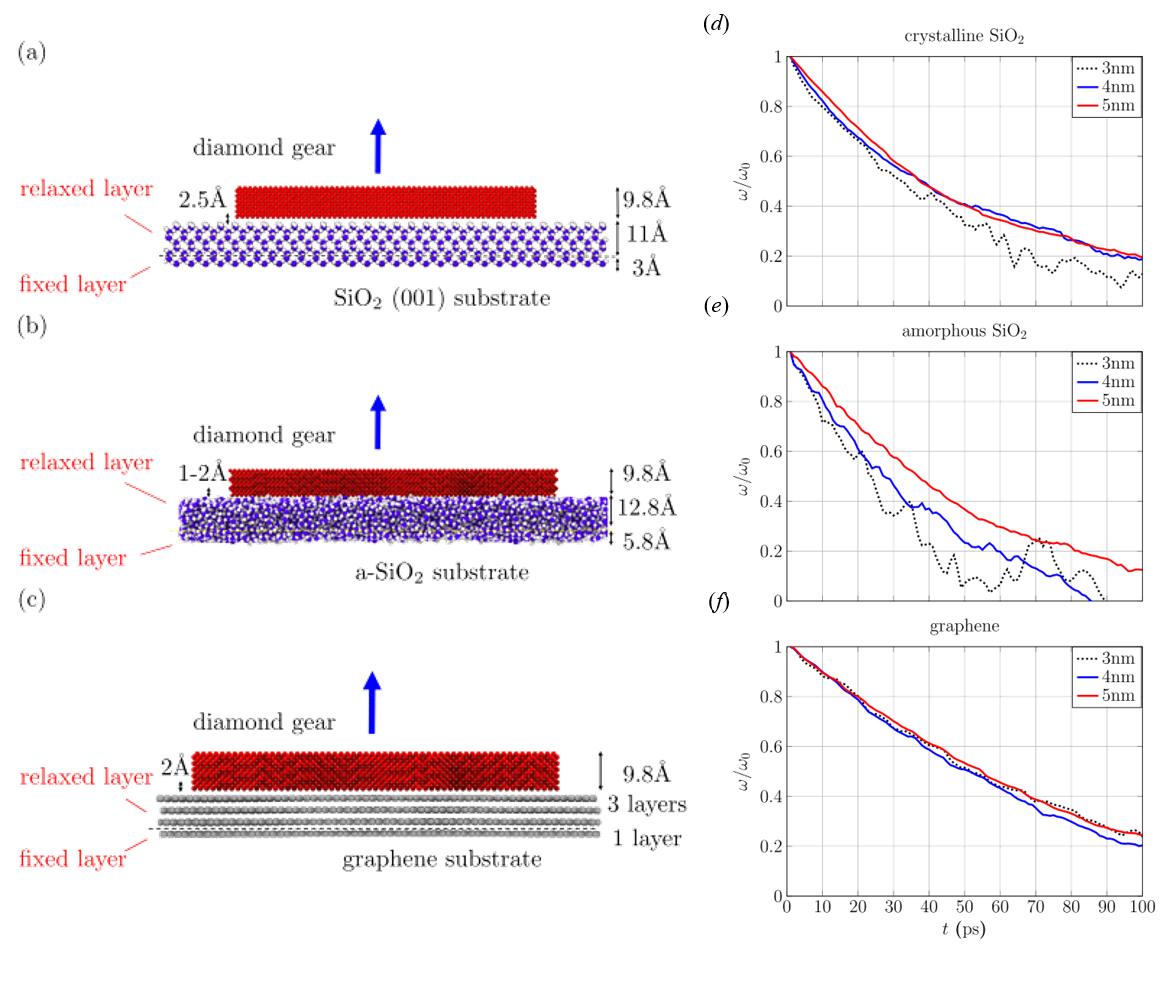}
 \caption{Scheme of the $5$ nm diamond-based solid-state gear on top of the (a) $\alpha$-cristobalite SiO$_2$ with size 23.0 nm$\times$15.3 nm$\times$1.39 nm and gear-substrate distance 2.5\AA;(b) amorphous SiO$_2$ (28.2 nm$\times$16.9 nm$\times$1.86 nm) and gear-substrate distance $\sim 1-$2\AA{ }, and (c) 4-layer graphene (21.8 nm$\times$12.6 nm$\times$ 4 monolayers) and gear-substrate distance 2\AA. The normalized angular velocity of the diamond-based solid-state gear ( $\omega_0=0.2$ rad/s) for different gear radius $r=3,4$ and 5 nm is shown in the right column for (d) crystalline SiO$_2$, (e) amorphous SiO$_2$ and (f) 4-layer graphene substrates, respectively. {Reused from Ref.\cite{Lin2020a} with permission: copyright American Physical Society} }
 \label{Fig:substrate}
\end{figure}

\textit{Substrate dependence of the angular velocity relaxation.$-$} To better understand the influence of a specific substrate on the relaxation process, we carried out MD simulations over $100$ ps at $T=10$ K with initial angular velocity $\omega_0=0.2$ rad/ps on three different substrates: crystalline SiO$_2$, amorphous SiO$_2$, and graphene, see Fig.\ \ref{Fig:substrate} (a)-(c). Besides the different substrates, we have also considered gears with three different diameters: 3,4, and 5 nm. The results showing the relaxation of the angular velocity for different substrates and gear sizes are shown in Figs.\ \ref{Fig:substrate} (d), (e), and (f). It becomes clear from the figures that decreasing the gear radius leads in general to an increase of the angular velocity fluctuations, especially for $\alpha$-cristobalite and amorphous SiO$_2$, although the effect is less pronounced for the graphene substrate. The origin of this behavior is that a reduction in gear size implies a reduction of the rotational kinetic energy, so that the gear will be more sensitive to surface vibrations, entering the low average tangential velocity dissipation regime where Brownian behavior becomes dominant. A further analysis of the relaxation times $\tau_{rel}$\cite{Lin2020a} shows that for crystalline SiO$_2$ (and even more for graphene) $\tau_{rel}$ is approximately gear-size independent, while for amorphous SiO$_2$ $\tau_{rel}$ becomes smaller with decreasing gear size. This behavior is related to the decreasing degree of surface corrugation when going from amorphous SiO$_2$ to crystalline SiO$_2$ and to graphene. 

\begin{figure}[t]\centering
 \includegraphics[width=0.99\textwidth]{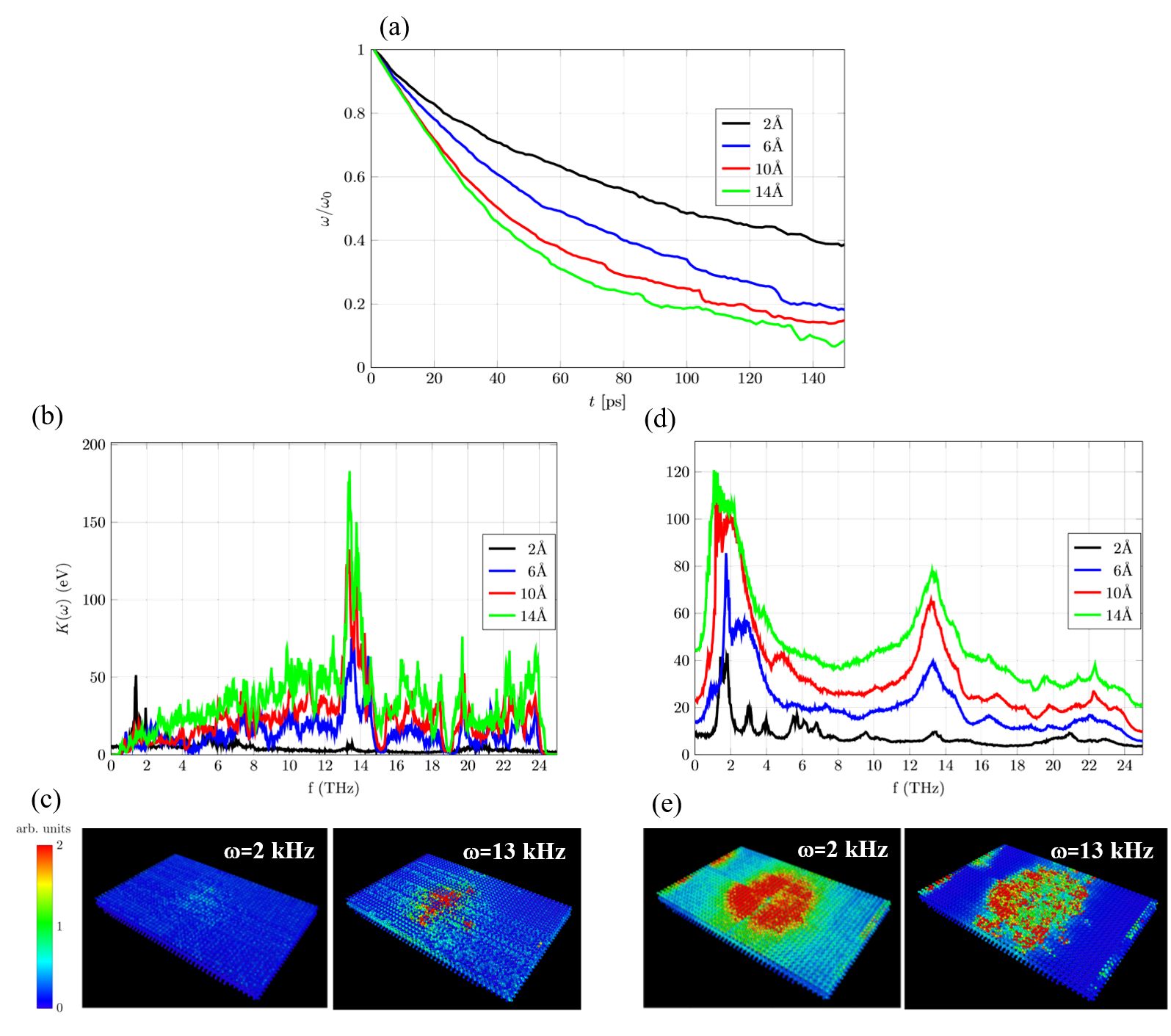}
 \caption{(a) Substrate thickness dependence of the angular velocity relaxation. The kinetic energy spectra of the substrate (b) without and (d) with gear rotation within $50$ ps. The heat maps in panels (c) and (e) present a real-space visualization of the kinetic energy spectrum $K_j(\nu_0)$ with contributions around the frequencies $\nu=2$ THz and $\nu=13$ THz (c) without and (e) with rotation of the solid-state gear. {Modified from Ref.\cite{Lin2020a} with permission: copyright American Physical Society}}
 \label{Fig:thickness_phonon}
\end{figure}

\textit{Scrutinizing surface phonon excitations.$-$} Although part of the rotational kinetic energy is dissipated into internal degrees of freedom of the gear\cite{Lin2020a}, the major dissipation channel is related to the excitation of surface vibrations. It is, therefore, interesting to see how the energy dissipation is modified when the thickness $d$ of the relaxed layer is changed (note that the overall thickness of the substrate in the simulations is always the same, only the thickness of the fixed layer is varied). The results for the crystalline SiO$_2$ substrate are shown in Fig.\ \ref{Fig:thickness_phonon} (a). One can see that with increasing $d$ the angular velocity relaxation time $\tau_{rel}$ becomes shorter (i.e., faster exponential decay). However, for larger $d$ there is a tendency of the decay rate to saturate. 

A deeper understanding of the energy dissipation process can be gained by analyzing the kinetic energy spectrum $K(\nu)$ \cite{Meshkov1997,Lee1993} as given by:
\begin{equation}
 K(\nu)=\sum_{i\in \text{R}}K_i(\nu)=\sum_{i\in \text{R}}\frac{1}{2}m_i|\mathbf{v}_i(\nu)|^2 \;,
\end{equation}
with $\mathbf{v}_i(\nu)$ being the Fourier transform of the velocity of the $i^{\text{th}}$ atom in the relaxed layer (R) and $\nu$ is the frequency. 
To identify the features arising purely from rotational dissipation, we show in Figs.\ \ref{Fig:thickness_phonon} (b) and (d) the kinetic energy spectra for the cases without and with gear rotation, respectively. 
For the case without rotation, Fig.\ \ref{Fig:thickness_phonon} (b), as $d$ increases a prominent spectral feature around $13$ THz emerges, which is related to the well-known van-Hove singularity in the phonon density of states of crystalline SiO$_2$\cite{Wehinger2015}. 

When the gear rotational motion is allowed, then two main features can be observed in the kinetic energy spectral function, see Fig.\ \ref{Fig:thickness_phonon} (e). First of all, the spectral feature at $\omega=13$ THz gets strongly broadened (its lifetime is shorter) and loses intensity with decreasing $d$. This latter effect indicates that this peak is a bulk property rather than a surface property. The former effect hints at the increased energy dissipation into the substrate via phonon excitation. Secondly, a new spectral feature appears at $\omega\sim 1-2$ THz. The corresponding peak intensity increases with increasing $d$, but the trend weakens for $d>10$ \AA{}, suggesting that only surface phonons are excited.

The bottom panels Fig.\ \ref{Fig:thickness_phonon} (c) and Fig.\ \ref{Fig:thickness_phonon} (d) provide a visualization of the spatial distribution of the two previously discussed spectral features at $\nu=2$ THz and $\nu=13$ THz. The calculation has been done by applying a band filter to the kinetic energy spectrum with bandwidth $\Gamma=1$ THz \cite{Lin2020a}.

Fig.\ \ref{Fig:thickness_phonon} (c) shows that without gear rotation more atoms are excited at $13$ THz than at $\omega=2$ THz, which is consistent with the features found in the kinetic energy spectrum. In contrast, upon allowing for gear rotation (Fig.\ \ref{Fig:thickness_phonon} (e)) the mode at $\omega=2$ THz has been strongly excited and localizes below the gear (the red region around the center), but also for $\omega=13$ THz, there is also a visible pattern, which is apparently due to the gear rotation. The conclusion from this analysis is that the low-frequency excitation is largely related to surface phonon modes excited by the gear rotation.



\subsection{\label{sec:graphene} Graphene nanodisk-molecule gear interaction}
\begin{figure}[t]
\centering
 \includegraphics[width=0.99\textwidth]{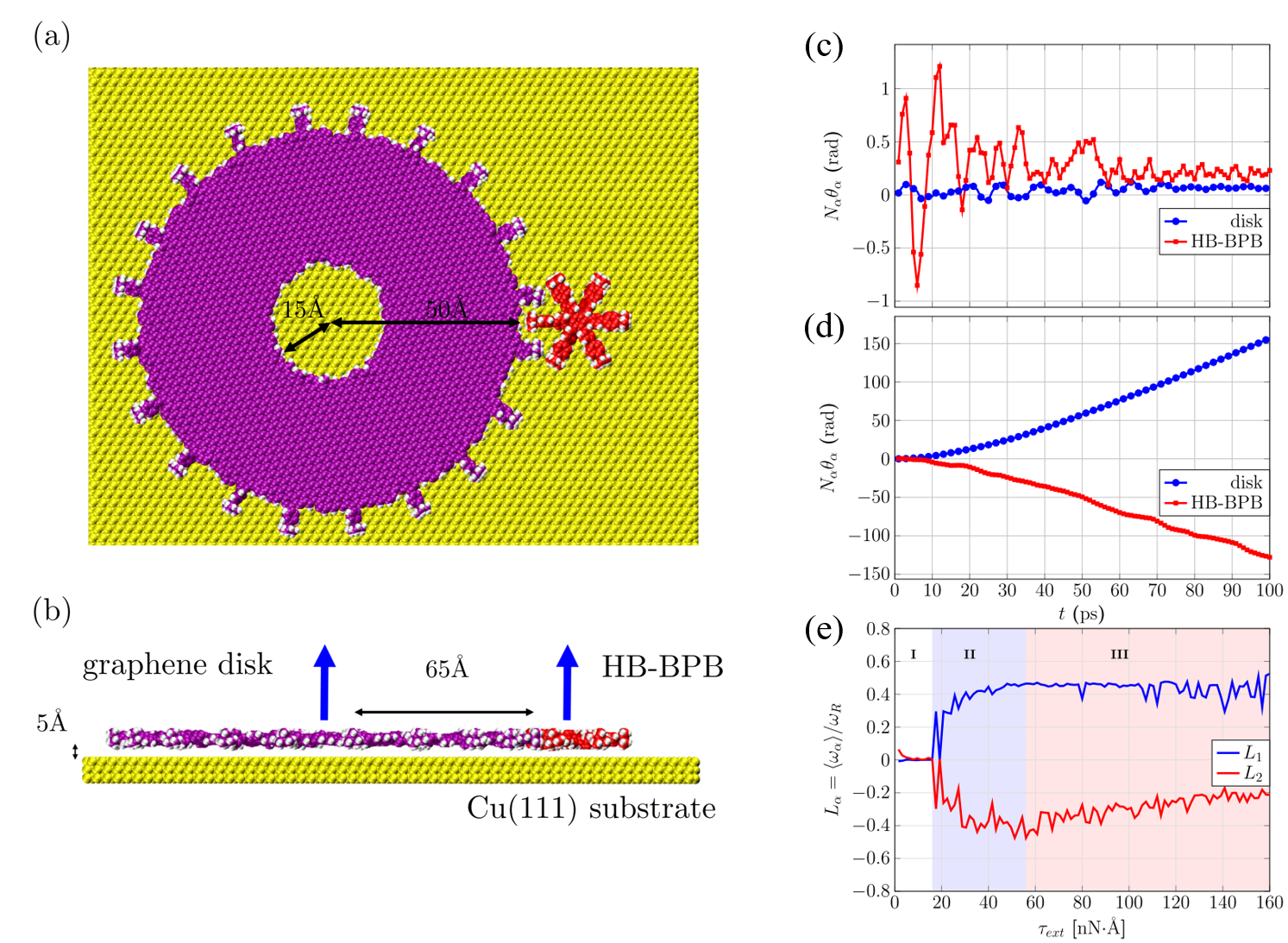}
 \caption{A schematic illustration of the (a) top view (b) side view for a \textit{tert}-butylbiphenyl functionalized graphene disk with radius 10 nm interacting with a hexa-\textit{tert}-butylbiphenylbenzene molecule gear on a Cu(111) surface. Panels (c) and (d) show the time dependence of the angular displacement $\theta_{\alpha}$ multiplied with the number of teeth $N_{\alpha}$ ($N_1=20$ for the graphene disk and $N_2=6$ for the HB-BPB molecule) for two applied external torques: (c) $\tau_{ext}=12.8$ nN$\cdot$\AA{} and (d) $\tau_{ext}=48$ nN$\cdot$\AA{}. Panel (e) displays the locking coefficient $L_{1,2}$ as a function of the applied external torque in the range 0 to 160 nN$\cdot$\AA{}. Regions I, II and III represent underdriving, driving and overdriving phases, respectively. {Modified from Ref.\cite{lin2021nanographene} with permission: copyright Institute of Physics}}
 \label{Fig:setup}
\end{figure}

As discussed in the Introduction, another interesting problem for the development and down scaling of gear systems is the possibility to couple gears with quite different sizes and to explore the efficiency of rotational transmission and energy dissipation. We have therefore studied in \cite{lin2021nanographene} a model system consisting of a graphene nanodisk and a single molecule gear, both adsorbed on a Cu(111) surface, see Fig.~\ref{Fig:setup} a). The molecule gear is a hexa-\textit{tert} -butylphenylbenzene (HB-BPB) molecule with 6 \textit{tert}-butylphenyl teeth. Further structural details as well as the rationale for choosing this molecule gear are provided in the next paragraphs.

 In what follows, we will use a simple way to characterize the mechanical transmission of rotational motion between two gears by defining a locking coefficient as\cite{Lin2020}:
\begin{equation}
 L_{\alpha}=\frac{\langle \omega_{\alpha} \rangle}{\omega_{R\alpha}}\;,
 \label{eq:Locking}
\end{equation}
where $\langle \omega_{\alpha}\rangle$ is the average angular velocity of gear $\alpha$ and $\omega_{R\alpha}$ is the terminal angular velocity of a rigid-body corresponding to this gear. For perfectly interlocked rigid gears, the coefficients $L_{\alpha}$ must be equal to unity. For soft gears or imperfect transmission, $L_{\alpha}$ will in general be much smaller than unity, since there will be dissipation into internal degrees of freedom.

\begin{figure}[t]
 \centering
 \includegraphics[width=\textwidth]{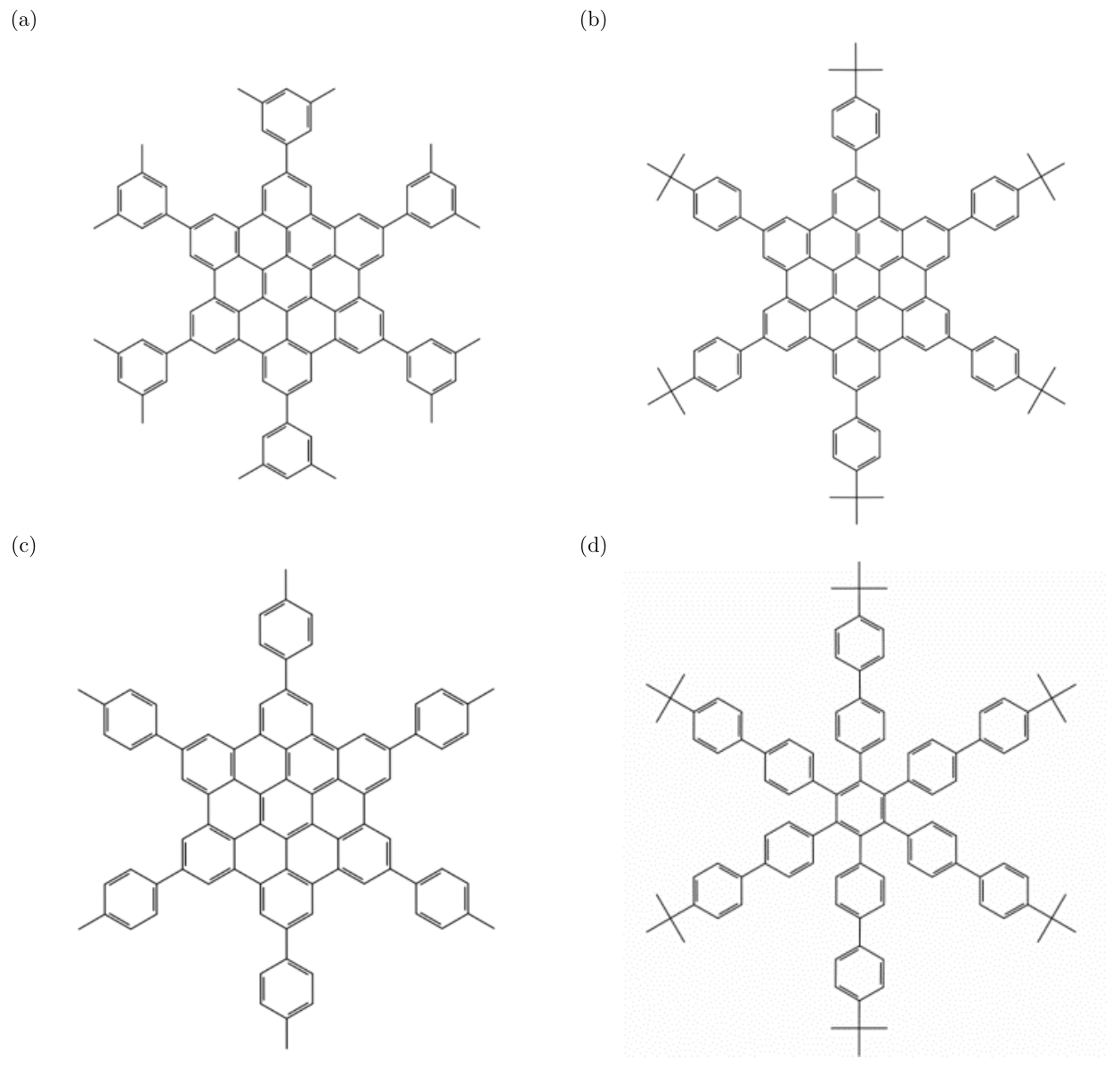}
 \caption{Some possible candidates for the molecule gear in the heterogeneous grapphene nanodisk-molecule gear system.}
 \label{Fig_MG:application} 
\end{figure}

\textit{Which molecule gear?$-$} Our choice of the molecule gear has been based on the test of few similar candidates and on their capability to efficiently transfer angular momentum in a molecule gear pair. In Fig.~\ref{Fig_MG:application}\, we show these molecules, which differ both in the compactness of their cores (alkyl groups or coronene cores) as well as on the shape of the teeth (adding one more phenyl ring). 
%

\begin{figure}[t]
 \centering
 \includegraphics[width=\textwidth]{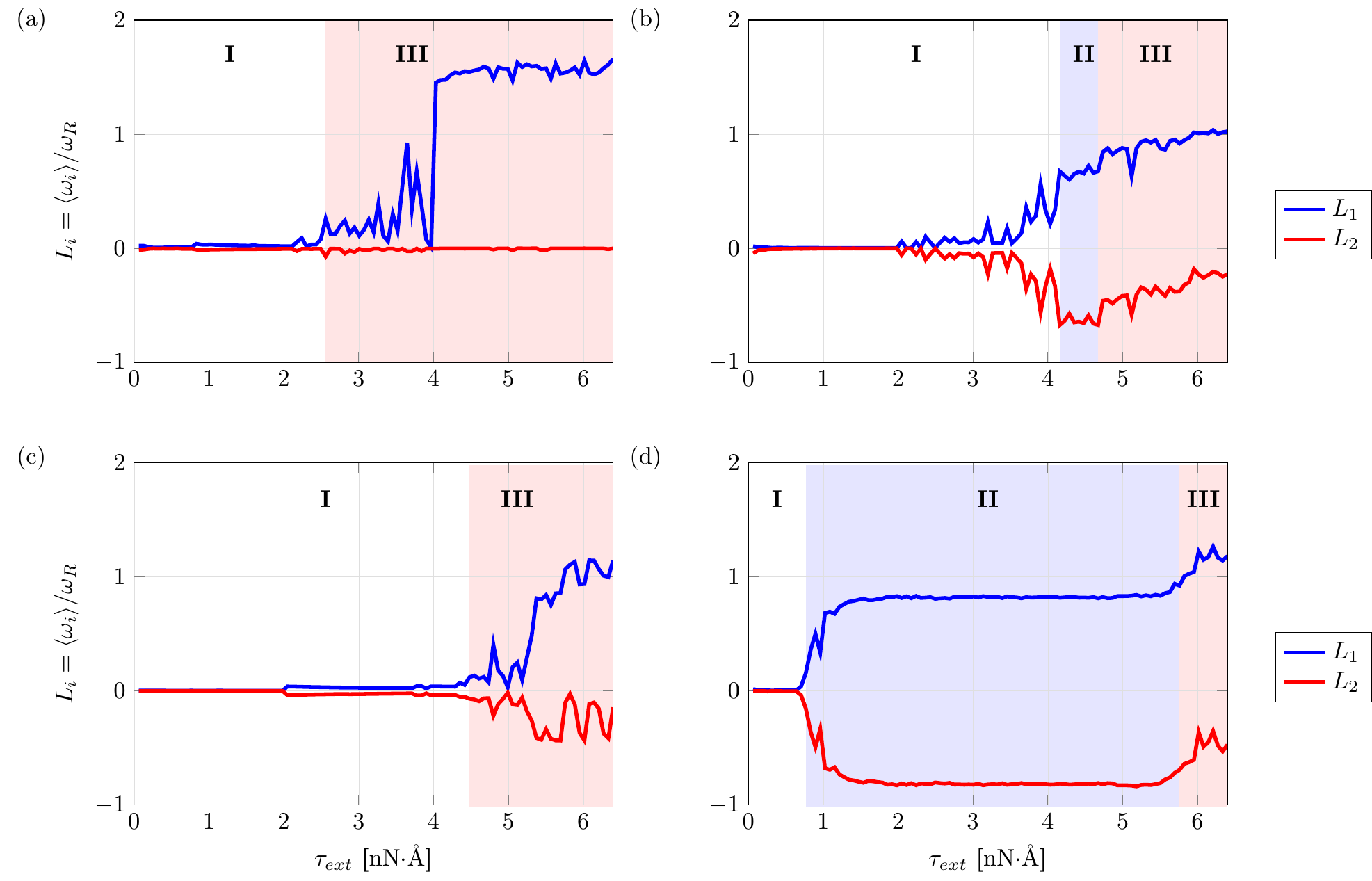}
 \caption{Locking coefficients diagrams for two gears in a train of the molecules depicted in Fig.~\ref{Fig_MG:application} with corresponding center-of-mass distances given by (a) $d_{\text{CM}}=2.25$ nm, (b) $d_{\text{CM}}=2.2$ nm (c) $d_{\text{CM}}=2$ nm and (d) $d_{\text{CM}}=2.2$ nm. Panels (a)-(d) correspond to the molecules in Fig.~\ref{Fig_MG:application}}
 \label{Fig_MG:application_locking} 
\end{figure}

To characterize the rotational transmission of the molecule gears we use the previously defined locking coefficient, Eq.~(3), and plot it as a function of an external torque applied on the first gear. As previously shown in Ref. \cite{Lin2020} for molecule gears, three different regimes can be identified (which do not necessarily appear simultaneously in each case), depending on the strength of the applied torque: (I) underdriving regime for which $|L_1|\approx |L_2|\approx 0$, i.e. the average angular velocities for both gears nearly vanish and there is no rotation; (II) driving regime with $0<|L_1|\approx |L_2|<1$, where transmission of rotation between the two gears takes place; (III) overdriving regime with $|L_1|>|L_2|$, where the driven gear cannot follow the driver. The goal is, therefore, to design a molecule gear where the driving regime is dominant over a broad range of applied torques. The corresponding phase diagrams for the four molecules (a) to (d) of Fig.~\ref{Fig_MG:application} are shown in Fig.~\ref{Fig_MG:application_locking}. 
%
%
For molecule (a), the distance between gears is $d_{\text{CM}}=2.25$ nm and only phases I and III were identified, so that no collective rotations take place. For molecule (b), the distance is $d_{\text{CM}}=2.2$ nm, and now all three phases appear, meaning that collective rotations are possible. For the molecule (c), $d_{\text{CM}}=2$ nm and we have again only regions I and III. Note that in the red region one can still see nonzero values for $L_2$ but the magnitude is relatively small compared with $L_1$. This implies that collective rotations are possible, but relative sliding between gears can happen as well. For molecule (d), $d_{\text{CM}}=2.2$ nm and one can immediately see that there is a wider region II compared to the other molecule gears, which means collective rotations are robust in terms of external torque. This is why we have chosen case (d), the hexa-\textit{tert}-butylbiphenylbenzene (HB-BPB) molecule, as our candidate to interface with the graphene nanodisk.

{\textit{Heterogeneous gears: graphene nanodisk-molecule gear.$-$}} 
After the previous discussion for choosing an appropriate molecule gear, we address now the heterogeneous system made of a graphene disk with a diameter of 10 nm and a hexa-\textit{tert} -butylphenylbenzene (HB-BPB) molecule with six \textit{tert}-butylphenyl teeth. For the graphene nanodisk we tested different edge profiles such as irregular edges (roughness), involute shape, and \textit{tert}-butylphenyl teeth. Only in the latter case a correlated rotation of both gears for clockwise and anti-clockwise directions of rotation was obtained. Therefore, this is the atomistic model chosen for the following simulations. 
We remark that the thickness of the graphene nanodisk is compatible with the molecule gear chemical structure and with its physisorption height on the used Cu(111) surface.
We have further found that using 20 \textit{tert}-butylpheny chemical groups provides an optimal number of teeth to ensure a smooth transmission of rotational motion. Other master diameters will certainly require to adapt again its edge chemical structure. 
There is also a central 1.5 nm hole in the master Fig.~\ref{Fig:setup} (a) representing the location of the rotational axle, which is not explicitly modelled in this study. The optimized center-of-mass separation between the master (graphene disk) and the HB-BPB molecule gear was found to be 6.5 nm.

We now apply an external torque $\tau_{ext}$ on the master and monitor the angular momentum transfer to the HB-BPB molecule. The time dependence of the product $N_{\alpha}\theta_{\alpha}(t)$ for a small $\tau_{ext}=12.8$ nN$\cdot$\AA{} and an intermediate $\tau_{ext}=48$ nN$\cdot$\AA{} torques are presented in Fig.~\ref{Fig:setup} (c) and (d, respectively, with $N_1= 20$ and $N_2 = 6$. For a small torque only small amplitude oscillations of both gears around their equilibrium conformations are found, since $\tau_{ext}$ is not large enough to overcome the surface friction of the master on the substrate and to allow the molecule gear to overcome the energy barrier of the potential energy surface on Cu(111) at the simulation temperature of $T$= 10 K. For intermediate $\tau_{ext}$, on the contrary, perfectly interlocked rotation is obtained, which satisfies the classical condition $N_1\theta_1+N_2\theta_2=0$\cite{MacKinnon2002}. 
In Fig.~\ref{Fig:setup} (e) we plot the locking coefficient as a function of the applied torque using $\omega_{R1}=\frac{\tau_{ext}}{\gamma_1}$ with $\gamma_1=2.83\times 10^{-29}$ kg$\cdot$m$^2\cdot$s$^{-1}$ (extracted from $\tau_{ext}/\langle\omega_{1}\rangle$ in the same MD simulation) being the damping coefficient corresponding to a rigid disk on the Cu surface. For $\omega_{R2}$, we then obtain $\omega_{R2}=-\frac{N_1\omega_1}{N_2}$. Interestingly, also for this heterogeneous two-gear system, we can still identify three typical different regimes (highlighted with different colors). In the (white) region I ($\tau_{ext}<18$ nN$\cdot$\AA{}) the average angular velocities for both gears nearly vanish, and there is no rotation (underdriving regime). In the (blue) region II ($18<\tau_{ext}<58$ nN$\cdot$\AA{}) transmission of rotation between the two gears takes place (driving regime). Finally, in the (red) region III ($\tau_{ext}>58$ nN$\cdot$\AA{}) the molecule gear cannot follow the master rotation (overdriving regime). Therefore, there is a relevant window in the applied torque (driving regime) where the interlocking of the two gears is largely favoured and the rotational transmission is optimal. 

\begin{figure}[t]\centering
 \includegraphics[width=0.99\textwidth]{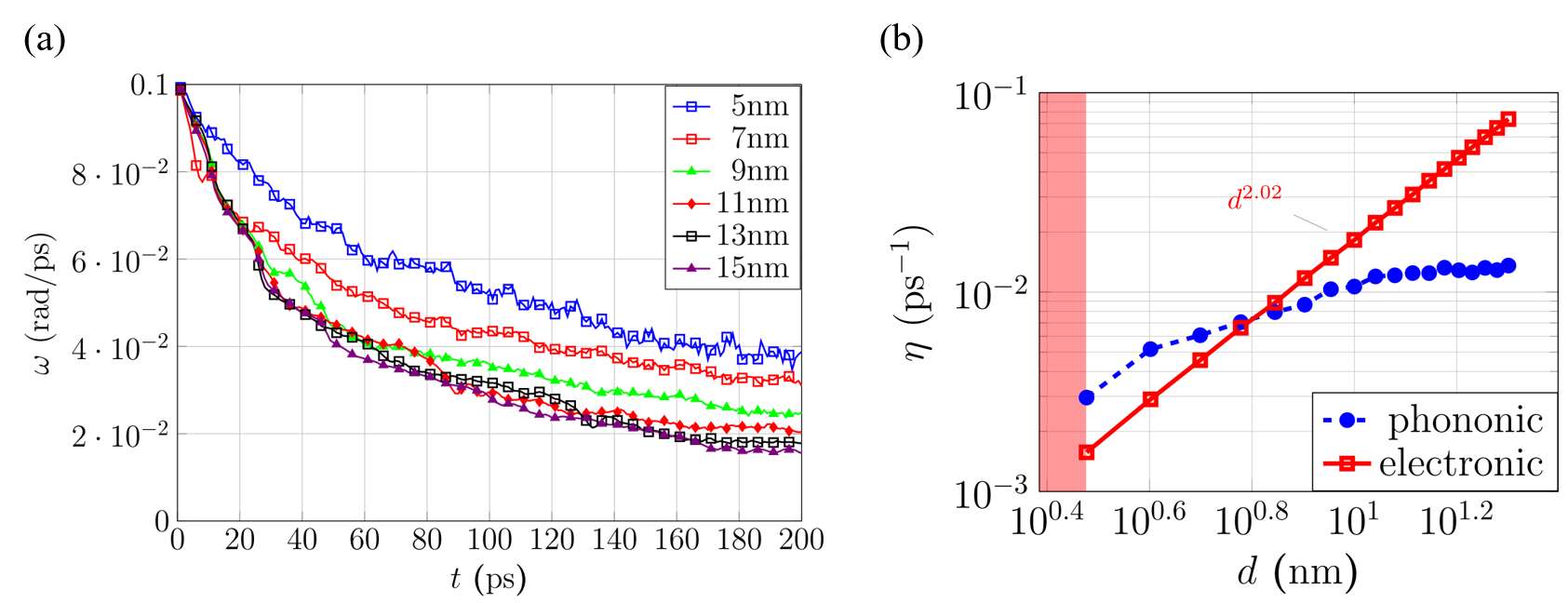}
 \caption{(a) Viscous dissipation for graphene disks with initial angular velocity $\omega_0=0.1$ rad/ps on Cu(111) within 200 ps for different gear diameters ranging from 5 to 15 nm. (d) Nanodisk diameter dependence of the rotational friction coefficient $\eta=1/\tau$ due to the phononic contribution (blue) and the electronic contribution according to Eq.~\eqref{eq:I} (red) for disk diameters $d$ from 3 to 20 nm. The red region indicates that below 3 nm the motion becomes oscillatory due to confinement by rotational barriers. {Modified from Ref.\cite{lin2021nanographene} with permission: copyright Institute of Physics}}
 \label{Fig:omega_relaxation_5_15nm}
\end{figure}

\textit{Nanodisk-molecule gear: Phononic dissipation channel.$-$} In the previously described situation the main source of energy dissipation (friction), which requires the application of a large torque, is the interaction between the master and the Cu(111) surface. To address in more detail this issue, we have considered graphene nanodisks with different diameters adsorbed on the Cu(111) surface, but without including the molecule gear. Then, by applying an initial angular velocity $\omega_0$ we allow the graphene disk to slow down and eventually stop its rotation as a result of frictional dissipation to the substrate. The estimated tangential velocities indicate that the system is in the viscous dissipation regime\cite{Guerra2010}. Hence, by plotting the time-dependent angular velocity as a function of simulation time for the different diameters, we can fit each curve to a single exponential function and can extract the corresponding relaxation times. The function $\omega(t)$ is shown in Fig.~\ref{Fig:omega_relaxation_5_15nm} (a). As it may be expected, the relaxation time is longer for small diameters than for larger diameters as a result of the increased contact area with the surface. 
%
%

The obtained inverse relaxation times $\tau^{-1}_{rel}$, related to the friction coefficient $\eta=1/\tau_{rel}$, are now displayed as a function of the nanodisk diameter as the blue curve in Fig.~\ref{Fig:omega_relaxation_5_15nm} (b). As seen in the figure, the friction increases with increasing diameter, but it becomes weakly dependent on $d$ for values larger than 10 nm. Notice also that for $d<3$ nm (highlighted by the vertical window in light red in Fig.~\ref{Fig:omega_relaxation_5_15nm} (b)), the nanodisks no longer rotate but oscillate in a Brownian like motion. 
The reason for this transition is that the potential energy barrier height for rotation becomes in this domain larger that the initially imposed rotational kinetic energy, so that only fluctuations around equilibrium positions are feasible. 

\textit{Nanodisk-molecule gear: Electronic dissipation channel.$-$} Although our classical MD simulations do not allow to directly address dissipation processes involving the electronic system, we can try to obtain estimates of it by using mathematical relations available from quantum mechanical studies of model systems. In the case of van der Waals related electronic friction, estimates from second-order perturbation theory using a Jellium model were obtained in Refs.\cite{Persson2000, Persson1995}:

\begin{equation}
 \eta_{el}=1/\tau_{el}=\frac{e^2}{ \hslash a_0}\frac{\left[k_F^3\alpha(0)\right]^2}{\left(k_Fz_0\right)^{10}}\frac{m_e}{M}\frac{\omega_F}{\omega_p}k_Fz_0I(z_0,r_s).\label{eq:I}
\end{equation}

Here, $e$ is the elementary charge, $\hslash$ is the reduced Planck constant, $a_0$ the Bohr radius, $\alpha(0)$ the static polarizability of the adsorbate, $\omega_p$ the plasma frequency of the metallic substrate, $k_F$ the Fermi wave vector, $\omega_F$ the Fermi frequency, $m_e$ the electron mass, $M$ the mass of the graphene nanodisk and $I$ is the parallel frictional integral as a function of the electron gas radius $r_s$ and the distance $z_0$ between the graphene nanodisk and the substrate ($z_0=3.1$ \AA{} in our case). For Cu(111), $\omega_p=1.33\times 10^{16}$ rad/s \cite{Ordal1985}, $\omega_F=1.06\times 10^{16}$ rad/s \cite{Ashcroft1976}, $k_F=1.36\times 10^{10}$ A$^{-1}$ \cite{Ashcroft1976} and $I(z_0,r_s)\approx7$ \cite{Persson2000}($r_s=2.67$ for Cu). For a monolayer graphene sheet, $\alpha(0)=0.949$ \AA{}$^{3}$ per unit cell\cite{Kumar2016}. To estimate $\alpha(0)$ for a graphene nanodisk, we assume that this disk is large enough such that $\alpha(0)\approx 0.949\times N_c/2$ (one unit cell contains two carbon atoms) where $N_c$ is the number of carbon atoms in the graphene nanodisk. Using these parameter values, we can estimate the electronic contribution to the frictional process as a function of the nanodisk diameter. This is shown as the dark red curve in Fig.~\ref{Fig:omega_relaxation_5_15nm} (b). According to Eq.~\eqref{eq:I}, the electronic friction coefficient is roughly proportional to $d^2$ since the increase of polarizability is proportional to the number of carbon atoms and it grows faster than the increase of the mass of a graphene nanodisk ($\alpha(0)^2/M\propto d^2$). As it becomes clear from the figure, the electronic contribution to friction is dominant for larger disks whereas for smaller diameters the phononic dissipation channel dominates the process. As a consequence, a large (saturating phononic dissipation) graphene nanodisk of higher polarizability together with a substrate below the critical superconducting transition temperature (for suppressing the electronic dissipation channel) could provide an efficient starting setup where to build our modelled heterogeneous gear system. 
 
\begin{figure}[t]
 \centering
 \includegraphics[width=\textwidth]{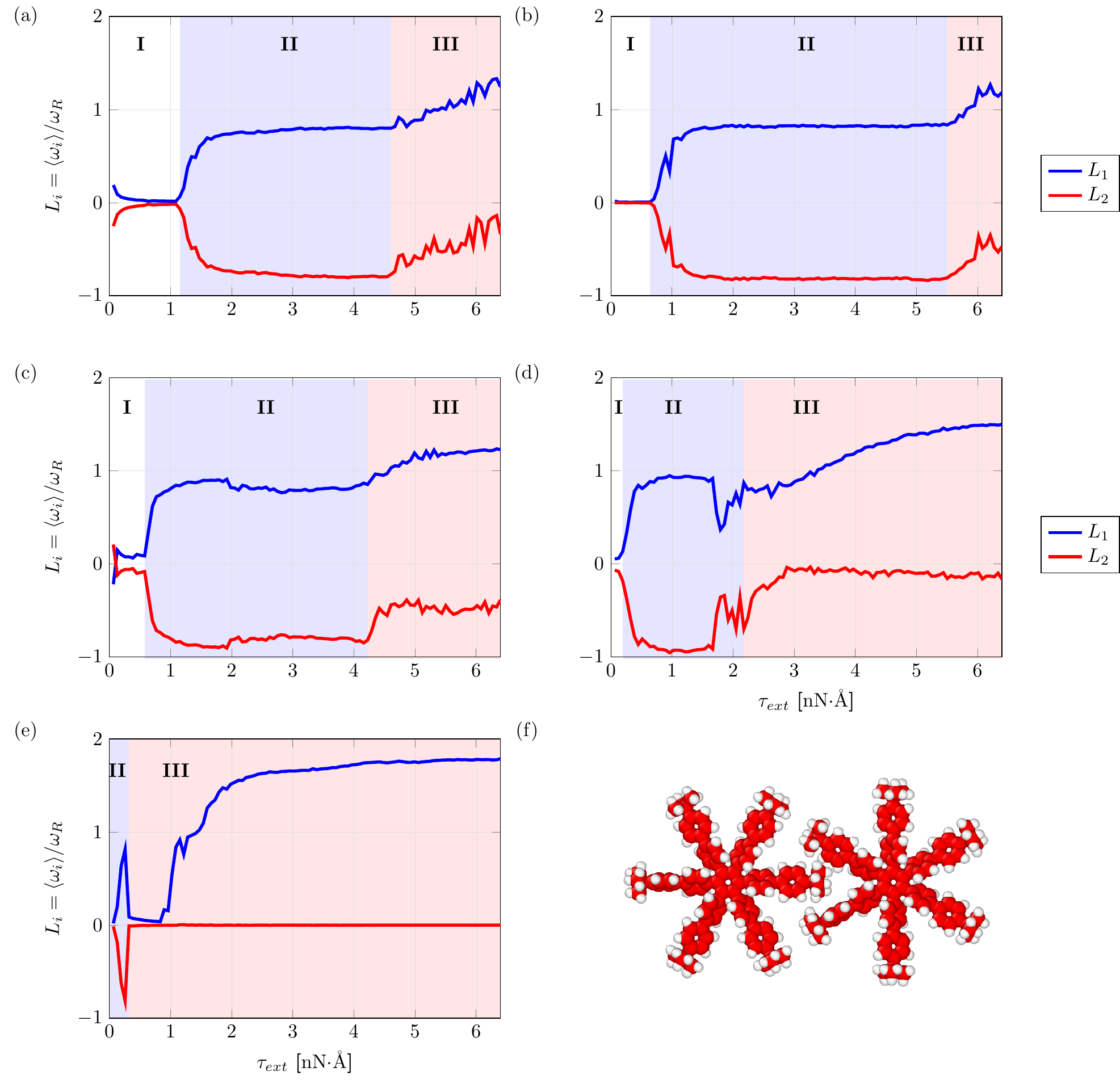}
 \caption{Locking coefficients diagrams for two hexa-\text{t}-butylbiphenylbenzene (HB-BPB) gears in a train with center-of-mass distance (a) $d_{\text{CM}}=2.1$ nm, (b) $d_{\text{CM}}=2.2$ nm, (c) $d_{\text{CM}}=2.3$ nm, (d) $d_{\text{CM}}=2.4$ nm and (e) $d_{\text{CM}}=2.5$. (f) Scheme of two HB-BPB gears.}
 \label{Fig_MG:Lp_distance} 
\end{figure}

\begin{figure}[t]
 \centering
 \includegraphics[width=\textwidth]{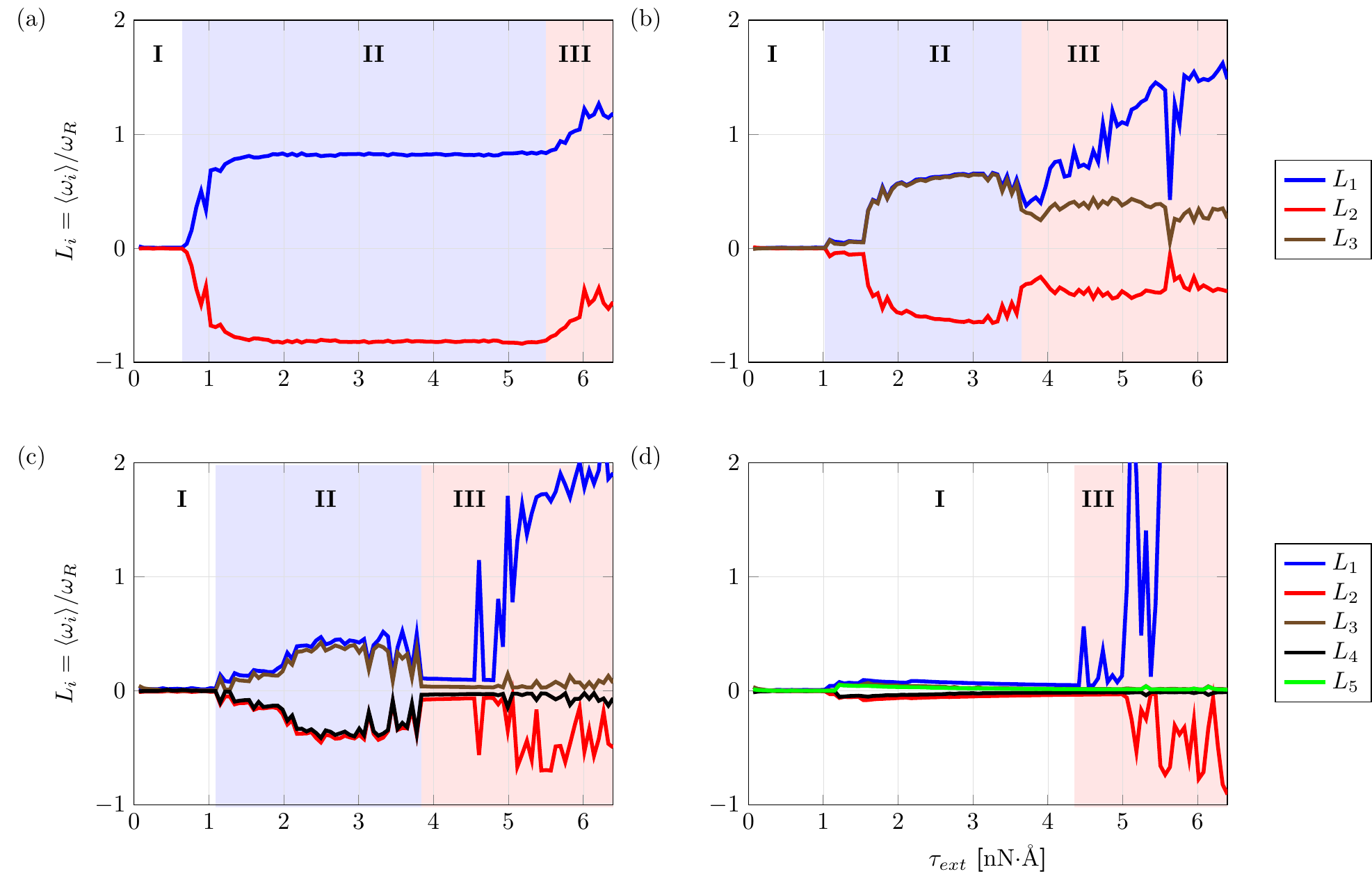}
 \caption{Locking coefficients diagrams for $N$ hexa-\text{t}-butylbiphenylbenzene (HB-BPB) gears in a train with center-of-mass distance $d_{\text{CM}}=2.2$ nm: (a) $N=2$ (b) $N=3$ (c) $N=4$ and (d) $N=5$.}
 \label{Fig_MG:Lp_Ngears} 
\end{figure}

\textit{Molecule gear trains based on HB-BPB: How efficient is the transmission of rotational motion?$-$} The graphene nanodisk-molecule gear system can be considered to form the driving part of a molecule gear train with additional HB-BPB molecules. It is therefore of interest to look closely at the critical inter-gear separations in order to optimize the rotational transmission. 
In Fig.~\ref{Fig_MG:Lp_distance}, we show the corresponding locking coefficients of two HB-BPB gears (see also the scheme in Fig.~\ref{Fig_MG:Lp_distance} (f)) for different center-of-mass separations (a) $d_{\text{CM}}=2.1$ nm, (b) $2.2$ nm, (c) $2.3$ nm, (d) $2.4$ nm and (e) $2.5$ nm, respectively. One can see that gears with $d_{\text{CM}}=2.2$ nm give the optimal result for rotational transmission, since there is a wide range of applied torques corresponding to a driving phase (region II). This is also the separation used in Fig.~\ref{Fig_MG:application_locking}. For $d_{\text{CM}}=2.4$ nm, on the contrary, the driving region II is already considerably narrowed. For $d_{\text{CM}}=2.5$ nm, there is largely only overdriving phase.
%
%
For this optimal separation $d_{\text{CM}}=2.2$ nm, we have then added additional gears to the train to see over how many gears rotational transmission is still feasible. In Fig.~\ref{Fig_MG:Lp_Ngears} we show the locking coefficients for $N=2,3,4,5$ HB-BPB gears. 
%
%
First of all, we see that adding gears leads to a reduction in the values of the locking coefficients within the driving phase. If the locking coefficients decrease, this suggests additional energy dissipation during the rotational process. In fact, with increasing number of gears, a more detailed analysis of the MD simulations showed that collective rotations gradually become a set of internal torsional motions propagating among gears. In Fig.~\ref{Fig_MG:Lp_Ngears} (b), we can see that we have regions I, II, and III. Therefore, collective rotations for three gears are possible. On the other hand, in region III, we found that the locking coefficients satisfy $L_1>|L_2|=|L_3|$. This means that one overdrives the first gear but the second and third gears are still interlocked and can sometimes still follow the first gear. For four gears (see Fig.~\ref{Fig_MG:Lp_Ngears} (c)), collective rotations happen as well but in region III the locking coefficients satisfy $L_1>|L_2|\gg|L_3|\approx|L_4|\approx 0$, meaning that the first gear is overdriven, but only the second gear can sometimes follow and the others stay still. For five gears (in Fig.~\ref{Fig_MG:Lp_Ngears} (d)), region II completely vanishes, so that collective rotations are fully blocked. In region III, we have a similar situation as in the four gears case: $L_1\gg|L_2|\gg|L_3|\approx|L_4|\approx 0$, so that the first gear rotates very fast and only the second gear can follow at some point. Note also that in region I, we found very small nonzero values for all locking coefficients: $|L_1|\approx|L_2|\approx|L_3|\approx|L_4|\approx|L_5|\neq 0$. This is related to the fact that internal torsional motions are propagating from the first gear to the remaining gears, but the whole process is very slow and gets stuck eventually. Hence, the average angular velocity for all gears is nonzero but very small.

These results suggest that transmission of rotational motion may work up to $N=4$ gears, but one needs to further investigate this issue by including e.g. a substrate and analyzing which dissipation channels become efficient in this case. The problem can be, however, very difficult since not only phononic dissipation is expected to occur, but also electronic friction, requiring a quantum mechanical treatment. 

\section{\label{sec:Conclusion}Conclusions and Outlook}
In conclusion, we have shown how atomistic classical Molecular Dynamics simulations can contribute to shed light on different issues related to the dynamics of nanoscale gears. In the first part of this Chapter we demonstrated that the regime of viscous dissipation can emerge out of van-der-Waals interactions between gears and substrates. The fitted exponential decay of the angular velocity allowed, additionally, to obtain an estimate of the typical relaxation time scales associated with the process. Moreover, by changing the substrates and the gear sizes, we found out that the relaxation time monotonically decreases with decreasing gear size for amorphous SiO$_2$, but it is weakly dependent on size for crystalline SiO$_2$ and nearly size-independent for graphene, a result which correlates well with the differences in the degree of atomic-scale corrugation of these three substrates. We also identified the energy transfer between gear and substrate as the main energy dissipation channel involving the excitation of surface vibrational modes. 

In the second part of the Chapter, we investigated a heterogeneous pair of gears, consisting of a solid-state graphene nanodisk whose edge was regularly functionalized with chemical groups acting as atomic scale teeth. The latter were able to efficiently mediate the transfer of angular momentum to a single molecule gear upon the application of an external torque. Similar to previous studies on molecule gear trains\cite{Lin2019a} we could identify underdriving, driving and overdriving phases as a function of the applied torque. Moreover, we estimated the contribution of phononic and electronic energy dissipation channels leading eventually to a damping of the rotational motion of the master (graphene nanodisk). Our results support the use of superconducting surfaces in order to suppress electronic friction contributions\cite{WeiHyo2019}. We also showed that the locking coefficient is a reasonable descriptor not only to classify the different driving regimes but also to help in the selection of appropriate molecule gear candidates.


In relation to frictional effects, future research lines will need to address in more detail rotational energy dissipation at the single molecule gear level, which will involve dealing not only with excited vibrational modes on the substrate, but also (for metallic substrates) with electronic dissipation channels. This will require the use of more advanced first-principle methodologies or reactive force-fields (like ReaxFF)\cite{Chenoweth2008} (if MD simulations are the goal). From the more phenomenological perspective, it would be interesting to develop effective models to describe rotational damping along lines similar to the formulation of the Caldeira-Leggett and Rubin models\cite{Weiss2012}, which are based in tracing out bath degrees of freedom and its inclusion in effective memory kernels, leading to a very rich physics. 

One important point not covered in this short review and, in general, not well-understood is the mechanism for triggering the rotation of molecule gears. While the "mechanics" of pushing a molecule gear with an STM tip may be understood with the help of \textit{ab initio} approaches and MD simulations, the excitation of a collective rotation via voltage pulses is considerably less clear\cite{Lin2019,PES}. For instance, what breaks the rotational symmetry? Which molecular orbitals are involved? How the inelastic electronic excitation can be "converted" into the excitation of a collective coordinate leading to a concerted motion (rotation) of the molecular frame? The theoretical description of these effects may require the use of quantum master equation approaches or mixed quantum-classical methodologies.

\begin{acknowledgement}
We would like to thank C. Joachim, J.~Heinze, A.~Mendez, A.~Raptakis, T.~K{\"u}hne, D.~Bodesheim, S.~Kampmann, R.~Biele, D. Ryndyk, A.~Dianat, and F.~Moresco for very useful discussions and suggestions. This work has been supported by the International Max Planck Research School (IMPRS) for ``Many-Particle Systems in Structured Environments'' and also by the European Union Horizon 2020 FET Open project "Mechanics with Molecules" (MEMO, grant nr.\ 766864).
\end{acknowledgement}


\end{document}